\documentclass[%
 reprint,
 superscriptaddress,preprintnumbers,
 nofootinbib,
 amssymb,
 aps,
]{revtex4-1}

\usepackage[utf8]{inputenc}
\usepackage{hyperref}
\usepackage[normalem]{ulem}
\usepackage{amsmath,amssymb,mathtools}
\usepackage{dsfont}
\usepackage{graphicx}  
\usepackage{braket}
\usepackage[table]{xcolor}
\usepackage{colortbl}
\usepackage{url}
\usepackage{extarrows}
\usepackage[normalem]{ulem}
\usepackage{slashed}
\usepackage[capitalise, english]{cleveref}
\usepackage[noindentafter]{titlesec} 

\allowdisplaybreaks

\usepackage{siunitx}
\usepackage{xspace}

    \makeatletter
    \def\CT@@do@color{%
      \global\let\CT@do@color\relax
            \@tempdima\wd\z@
            \advance\@tempdima\@tempdimb
            \advance\@tempdima\@tempdimc
    \advance\@tempdimb\tabcolsep
    \advance\@tempdimc\tabcolsep
    \advance\@tempdima2\tabcolsep
            \kern-\@tempdimb
            \leaders\vrule
                    \hskip\@tempdima\@plus  1fill
            \kern-\@tempdimc
            \hskip-\wd\z@ \@plus -1fill }
    \makeatother

\usepackage{accents}
\DeclareMathSymbol{\widetildesym}{\mathord}{largesymbols}{"65}

\titleformat*{\subsubsection}{\normalfont \small \bfseries \boldmath}
\renewcommand{\paragraph}[1]{\vspace{.3em} \indent {\bfseries \boldmath #1 ---}\xspace }
\makeatletter
    \renewcommand{\p@subsection}{}
    \renewcommand{\p@subsubsection}{}
\makeatother

\definecolor{red}{rgb}{0.6,.0706,.1373}
\definecolor{blue}{rgb}{0,0.396,0.741}
\newcommand\myshade{80}
\colorlet{mylinkcolor}{violet}
\colorlet{mycitecolor}{violet}
\colorlet{myurlcolor}{violet}

\hypersetup{
  linkcolor  = mylinkcolor!\myshade!black,
  citecolor  = mycitecolor!\myshade!black,
  urlcolor   = myurlcolor!\myshade!black,
  colorlinks = true
}
\newcommand{\eps}{\varepsilon}

\newcommand{\be}{\begin{equation}}
\newcommand{\ee}{\end{equation}}
\newcommand{\bea}{\begin{eqnarray}}
\newcommand{\eea}{\end{eqnarray}}

\newcommand{\win}{w_{1}}
\newcommand{\wout}{w_{2}}


\begin{document}


\title{
Renormalization of the primordial inflationary power spectra
}

\author{Silvia Pla}
\email{silvia.pla\_garcia@kcl.ac.uk}
\affiliation{Physics Department, King’s College London, Strand, London, WC2R 2LS, United Kingdom}

\author{Ben A. Stefanek}
\email{benjamin.stefanek@kcl.ac.uk}
\affiliation{Physics Department, King’s College London, Strand, London, WC2R 2LS, United Kingdom}

\preprint{KCL-PH-TH/2024-10}

\begin{abstract}
It has been suggested that the effects of renormalization significantly reduce the amplitude of the inflationary spectra at scales measurable in the cosmic microwave background. Via a gauge-invariant analysis, we compute the renormalized scalar and tensor power spectra and follow their evolution in an inflating universe that undergoes a transition to an FRW phase with a growing horizon.  For perturbations originating from Minkowski vacuum fluctuations, we show that the standard prediction for the spectra on superhorizon scales is a late-time attractor, while they are UV finite at all times. Our result is independent of the equation of state after inflation, showing that the standard prediction is fully robust.
\end{abstract}

\maketitle

\section{Introduction} 
\label{sec:intro}
The theory of inflation~\cite{Guth:1980zm,Starobinsky:1980te,Linde:1981mu} provides an elegant solution to the horizon and flatness problems, as well as an explanation for the homogeneity and isotropy of the universe. Soon after it was proposed, it was realized that the union of quantum theory and inflation predicted the generation of a nearly scale-free spectrum of primordial scalar and tensor fluctuations~\cite{Starobinsky:1979ty,Mukhanov:1981xt,Hawking:1982cz,Starobinsky:1982ee,Guth:1982ec,Linde:1982uu,Starobinsky:1983zz,Bardeen:1983qw,Halliwell:1984eu}. The former, observed as temperature anisotropies in the cosmic microwave background (CMB)~\cite{COBE:1992syq,WMAP:2003elm,WMAP:2010qai,Planck:2013pxb,Planck:2015fie,Planck:2018vyg}, act as the seeds for structure in the universe, while the latter leads to the prediction of a spectrum of relic gravitational waves carrying information from the earliest moments of the universe.

Currently, only an upper bound on the amplitude of the tensor perturbations exists, quantified in terms of the tensor-to-scalar ratio $r < 0.036~(95\%$ CL)~\cite{BICEP:2021xfz}. Among the models favored by current data~\cite{Planck:2018jri} are those featuring a spontaneously broken conformal symmetry, predicting $r \approx 12/N_{\rm CMB}^2 \gtrsim 10^{-3}$~\cite{Starobinsky:1980te,Bezrukov:2007ep,Kallosh:2013hoa,Kallosh:2013daa,Ferrara:2013rsa,Farakos:2013cqa,Mukhanov:2013tua}. This is within the future sensitivity on $r$, expected to improve by an order of magnitude in the next decade~\cite{CMB-S4:2016ple,Abazajian:2019eic,SimonsObservatory:2018koc}. This makes a detection of primordial gravitational waves possible in the near future, which would correspond to a measurement of the energy scale of inflation. It is therefore crucial to have firm theoretical predictions for the tensor and scalar spectra in order to translate CMB measurements into constraints on model parameters.

The power spectra are defined by expressing the basic object of interest, the equal-time two-point function, as an integral over modes in momentum space. The fact that this object has ultraviolet (UV) divergences in the coincidence  limit $x' \rightarrow x$ has led to the argument that it should be renormalized~\cite{Parker:2007ni,Agullo:2008ka, Agullo:2009vq,Agullo:2009zi,delRio:2014aua,Urakawa:2009xaa}.
In particular, it was suggested in Ref.~\cite{Parker:2007ni} that renormalization results in a significant reduction of the amplitude of the spectra at CMB scales. This idea was further investigated in~\cite{Agullo:2008ka, Agullo:2009vq,Agullo:2009zi,delRio:2014aua}, where non-standard relations between CMB measurements and inflationary model parameters were found.\footnote{For an application to PBH formation see, for example, Ref. \cite{Choudhury:2023rks}.} 
Specifically, they apply the method of \emph{adiabatic regularization}~\cite{Parker:1974qw,Fulling:1974zr,Fulling:1974pu,Birrell:1982ix,Mukhanov:2007zz,Parker:2009uva} in order to derive a set of local counterterms that render the spectrum UV finite. They obtain their final result by evaluating these counterterms on a mode-by-mode basis at the moment of horizon crossing during inflation, leading to a severe reduction in amplitude compared to the standard prediction. 

This result has sparked a vigorous debate that continues unresolved at the present time~\cite{Durrer:2009ii,Urakawa:2009xaa,Marozzi:2011da,Agullo:2011qg,delRio:2014aua,Bastero-Gil:2013nja,Wang:2015zfa,Markkanen:2017rvi,Corba:2022ugu,Ferreiro:2022ibf,Ferreiro:2023uvr}. 
Two key points in the debate can be summarized as follows: i) since the non-coincident two-point function is already finite in a distributional sense, some authors argue that renormalization is unnecessary~\cite{Finelli:2007fr} or ii) the adiabatic method fails in some way and should be modified, but there is no agreement on how it fails or what the modification should be. For example, some claim that adiabatic regularization is not applicable on superhorizon scales~\cite{Durrer:2009ii,Marozzi:2011da}, or that it should be modified by introducing new, arbitrary scales~\cite{Ferreiro:2022ibf,Ferreiro:2023uvr}. Further discussion regarding point i) can be found in Ref.~\cite{Agullo:2011qg,Woodard:2017zfq}.  Here, we will show that this issue is practically irrelevant by investigating the consequences of renormalization.

A third viewpoint, first advocated in Ref.~\cite{Urakawa:2009xaa}, is that there is no problem with the adiabatic method, but evaluating the counterterms at the moment of horizon crossing is not appropriate. This is because the counterterms continue to evolve after horizon crossing, unlike the comoving curvature perturbation which has a well-known freezing behavior. The authors of Refs.~\cite{Parker:2007ni,Agullo:2008ka,Agullo:2009vq,Agullo:2009zi,delRio:2014aua} argue for their choice claiming there is a quantum-to-classical transition soon after horizon crossing, leaving the analysis of what happens at later times for future investigation. However, while it is true that the perturbations begin to approximate a classical Gaussian random field with exponential accuracy soon after horizon crossing, this process happens dynamically; there is no need to put it ``by hand"~\cite{Polarski:1995jg}. Indeed, if one starts from a pure quantum state deep inside the horizon, as is usually assumed, it remains pure under the subsequent unitary evolution. From this point of view, there is nothing special that happens at the moment of horizon crossing.

Since the adiabatic method requires the derived counterterms to be subtracted for all times and scales, we believe the argument briefly sketched in 2009 by Ref.~\cite{Urakawa:2009xaa} to be correct. However, the existence of continued debate in the literature shows that the community has not been convinced.
Here we show that, if, as demanded by the adiabatic prescription and general covariance, the adiabatic counterterms are subtracted on a \emph{mode-by-mode basis at all times}, the standard prediction for the superhorizon power spectrum is a late-time attractor. This is true whether inflation is eternal or lasts for a finite period. In the latter case, a transition to a Friedmann-Robertson-Walker (FRW) universe with a growing horizon occurs, which must be accounted for to obtain the correct result for the power spectrum (which is measured only at late times). In particular, the problematic scale-invariant counterterm becomes exponentially suppressed a few e-folds after the end of inflation, while the scalar and tensor perturbations remain frozen outside the horizon. Nevertheless, the spectrum remains UV finite at all times. We show all of this explicitly by matching solutions to the mode equations at the transition. We emphasize that no modification of the adiabatic regularization method is required.  Our conclusions are independent of the equation of state after inflation; the only requirement is that the modes start from Minkowski quantum vacuum fluctuations. 

This letter is organized as follows. In~\cref{sec:two} we review the standard computation of quantum fluctuations during inflation and compute the renormalized scalar and tensor power spectra. We comment on the reduction in power that would be found by evaluating the counterterms during inflation. Then in~\cref{sec:three} we show how standard inflationary predictions are an attractor solution in \emph{any} post-inflationary FRW universe using an instant transition model supported by a full numerical solution. Finally, we discuss our results and offer our conclusions in~\cref{sec:conc}.

\section{Quantum fluctuations during inflation} 
\label{sec:two}
For the sake of defining variables, we begin by writing the perturbed FRW metric in the longitudinal gauge~\cite{Mukhanov:1990me}
\begin{equation}
ds^2 = a^2\big[ -(1+2\Phi) d\tau^2 + [ (1-2\Psi) \delta_{ij} + h_{ij}] dx^i dx^j \big] \,,
\end{equation}
where $a(\tau)$ is the scale factor and $\tau$ is the conformal time defined by $d\tau = dt/a$. The functions $\Phi$ and $\Psi$ are equal to first order in single-field inflation. It is important to work with gauge-invariant variables~\cite{Bardeen:1980kt,Mukhanov:1988jd,Mukhanov:1989rq,Mukhanov:1990me}. The tensor fluctuations $h_{ij}$ are chosen to satisfy the transverse-traceless condition $\partial_i h_{ij} = h_{\,i}^{i} = 0$, which yields two physical polarizations $h= h_{+ ,\times}$~\cite{Misner:1973prb}. For the scalars, one can define the gauge-invariant quantity $\zeta$~\cite{Mukhanov:1990me}
\begin{equation}
\zeta = \Psi + H \frac{\delta \phi}{\dot \phi} \,,
\end{equation}
where $\phi$ is the inflaton field with perturbation $\delta\phi$, $H = \dot{a}/a$ is the Hubble parameter, and dots indicate differentiation with respect to $t$.  Our goal is to compute the coincident two-point functions of $\zeta$ and $h$ and renormalize them using the adiabatic regularization technique (reviewed in~\cref{sec:ADR}).

To this end, we define the Mukhanov variables
\begin{align}
&v_s = z_s \zeta \,, \qquad z_s = a M_P \sqrt{2\eps}  \,,\\
&v_t = z_t h \,,  \,\qquad  z_t = a M_P / 2 \,,
\end{align}
where the subscripts $s,t$ refer to scalar and tensor, $\eps = -\dot{H}/H^2$, and $M_P$ is the reduced Planck scale. These variables are convenient since they allow the scalar and tensor perturbations to be described by the same action~\cite{Mukhanov:1990me,Lidsey:1995np}
\begin{equation}
\mathcal{S} =\frac{1}{2} \int d\tau d^{3} {\bf x} \left[(v')^2 - (\nabla v)^2 +\frac{z''}{z} v^2 \right] \,,
\label{eq:action}
\end{equation}
where we have dropped the $s,t$ subscripts, primes indicate differentiation with respect to $\tau$, and $z(\tau)$ contains all the information about the gravitational background.
This action is equivalent to that of a free scalar field in Minkowski spacetime with a time-dependent mass term $z''/z$, so it is suitable for quantization via the canonical procedure~\cite{Birrell:1982ix}. We expand the operator $\hat v$ in plane waves
\begin{equation}
\hat  v(\tau, {\bf x}) = \int \frac{d^3{\bf k}}{(2\pi)^{3/2}} \left[ v_k(\tau) \hat a_{\bf k} e^{i{\bf k\cdot x}} + v_k^{*}(\tau) \hat a^{\dagger}_{\bf k} e^{-i{\bf k\cdot x}} \right] \,,
\label{eq:PWE}
\end{equation}
where $\hat a_{\bf k}\ket{0} = 0$ and $v_k(\tau)$ satisfies
\begin{equation}
v''_k (\tau) + \left(k^2 - \frac{z''}{z} \right) v_k(\tau) = 0 \,,
\label{eq:MSeq}
\end{equation}
with the normalization condition $v_k v_k^{\prime*} - v_k^* v_k^{\prime} = i$ such that $[\hat a_{\bf k}, \hat a^{\dagger}_{\bf k'}] = \delta^{(3)}({\bf k-k'})$.
We would like to compute the coincident equal-time two-point function $\langle v^2 \rangle \equiv \bra{0} \hat v(\tau, {\bf x}) \hat v(\tau, {\bf x}) \ket{0}$. Using~\cref{eq:PWE}, it can be written as
\begin{equation}
\langle v^2 \rangle = \int_{0}^{\infty} \frac{dk}{k}\mathcal{P}_{v}(k,\tau)\,, \hspace{7mm} \mathcal{P}_{v}\equiv \frac{k^3}{2\pi^2} |v_k(\tau)|^2 \,,
\end{equation}
where $\mathcal{P}_{v}$ is the unregularized power spectrum of the Mukhanov variable. The spectra for $\zeta$ and $h$ are easily obtained via $\mathcal{P}_{\zeta} = \mathcal{P}_{v_{s}}/z^2_{s}$ and $\mathcal{P}_{h} = 2\mathcal{P}_{v_{t}}/z^2_{t}$.\footnote{The factor of 2 in the tensor case comes from summing over the two independent polarizations, namely $\mathcal{P}_{h} \propto |h_{+}|^2+ |h_{\times}|^2  $ .}
We will first calculate the power spectrum in an inflating universe defined by $0 \leq \eps <1$.\footnote{While $\zeta$ becomes a gauge degree of freedom in the exact de Sitter case $\eps =0$~\cite{Maldacena:2002vr}, the Mukhanov variable remains well defined.} We fix the initial condition for $v_k$ by assuming that all modes were well inside the horizon at early times $\tau\rightarrow -\infty$, where they began in vacuum of Minkowski quantum field theory~\cite{Bunch:1978yq} 
\begin{equation}
v_{k} (\tau\rightarrow -\infty ) \, \longrightarrow \, \frac{1}{\sqrt{2k}} e^{-i k \tau} \,.
\label{eq:vBC}
\end{equation}
To solve~\cref{eq:MSeq}, we need $z''/z$ for both scalars and tensors. To leading order in the slow-roll parameters, they are~\cite{Lidsey:1995np}
\begin{align}
\frac{z_s''}{z_s} \approx  \frac{2}{\tau^2}\left( 1+3\varepsilon-\frac{3 \eta }{2}  \right) &\equiv \frac{\nu_s^2 - 1/4}{\tau^2} \\
\frac{z_t''}{z_t} = \frac{a''}{a} \approx  \frac{2}{\tau^2}\left( 1+2\varepsilon  \right) &\equiv \frac{\nu_t^2 - 1/4}{\tau^2}
\end{align}
where $\eta = \eps - \dot{\eps} / (2H\eps)$, $\nu_s \approx 3/2 +2\eps -\eta$, and $\nu_t \approx 3/2 + \eps$. So in the limit of constant slow-roll parameters we just need to solve~\cref{eq:MSeq} for some real constant $\nu \geq 3/2$. In this case, the analytic solution satisfying the boundary condition~\cref{eq:vBC} is
\begin{equation} 
v_{k}(\tau) = \sqrt{\frac{\pi}{4k}}  e^{ i\frac{\pi}{4}(1+2\nu)} \sqrt{q} \, H_{\nu}^{(1)}(q) \,,
\label{eq:BDmodes}
\end{equation}
where $H_{\nu}^{(1)}$ is the Hankel function of the first kind and we have defined $q = -k\tau$. Then we have
\begin{equation}
\mathcal{P}_{v}(k,\tau) = \frac{k^2}{8\pi} |\sqrt{q} \,H_{\nu}^{(1)}(q)|^2 \,.
\end{equation}
The UV expansion $q \gg 1$ gives
\begin{equation}
\mathcal{P}_{v}(k,\tau) \, \xlongrightarrow[\text{UV}]{} \, \frac{k^2}{4\pi^2} + \frac{(\nu^2-1/4)}{8\pi^2\tau^2} + O\left(k^{-2}\right) \,,
\label{eq:vUV}
\end{equation}
which yields an infinite contribution to $\langle v^2 \rangle$, since the integral is both quadratically and logarithmically divergent in the UV. These are precisely the divergences we would like to cure. We define the regularized spectrum as 
\begin{equation}
\mathcal{P}_{v}^{\rm reg} \equiv \mathcal{P}_{v}(k,\tau) - \mathcal{P}^{\rm ct}_{v}(k,\tau)\,,
  \label{eq:regPS}
\end{equation}
where $\mathcal{P}^{\rm ct}_v$ contains the adiabatic counterterms defined in~\cref{sec:ADR}
\begin{equation}
  \mathcal{P}^{\rm ct}_{v}(k,\tau) =  \frac{k^2}{4\pi^2} + \frac{1}{8\pi^2} \frac{z''}{z}\,.
  \label{eq:ctSpec}
\end{equation}
Subtracting $\mathcal{P}^{\rm ct}_{v}(k,\tau)$ from~\cref{eq:vUV} leads to an exact cancellation of both UV divergent terms.
Due to the scale-invariant term $\propto z''/z$, the adiabatic subtractions also change the infrared (IR) part of $\mathcal{P}_{v}^{\rm reg}$. Since the IR behavior of $\mathcal{P}_{v}$ in general does not correspond to $\mathcal{P}^{\rm ct}_{v}$, the scale-invariant term appears to introduce a new IR divergence in $\langle v^2 \rangle$. However, as we show in~\cref{sec:ADR}, this IR divergence is spurious since it can be removed by a UV finite redefinition of $\mathcal{P}^{\rm ct}_{v}$. It is well known that renormalization schemes may differ by such finite parts.  Indeed, this is an alternative way to see that the physical spectrum cannot have IR distortions due to renormalization, a process which involves adding local counterterms. However, in the spirit of comparing with the original adiabatic method as defined in~\cite{Parker:2007ni}, we proceed here to show that a sensible answer is anyway obtained if one uses~\cref{eq:ctSpec}.

In the limit of constant slow-roll parameters, $z''/z = a''/a$ and the scale factor can be found by direct integration to be $a \propto \tau^{1/2 -\nu}$. It follows that $aH\tau = (1/2-\nu)$. Using these relations and expanding $\mathcal{P}_{v}$ in the IR limit $q \ll 1$ gives
\begin{align}
  \mathcal{P}_{v} (k,\tau) \xrightarrow[\text{IR}]{} \frac{a^2 H_{*}^2}{4\pi^2}\frac{2^{2\nu-3}\Gamma(\nu)^2}{\Gamma(3/2)^2}\left(\nu-\frac{1}{2}\right)^{1-2\nu} \left(\frac{k}{k_*} \right)^{3-2\nu} \,,
  \label{eq:inIRps}
\end{align}
where $H_*$ is the Hubble rate when a reference scale $k_* = a_* H_*$ crossed the horizon. Similarly, for the counterterms we find
\begin{equation}
  \mathcal{P}^{\rm ct}_{v}(k,\tau) \xrightarrow[\text{IR}]{}  \frac{a^2 H_{*}^2}{8\pi^2}\left(\frac{\nu+1/2}{\nu-1/2}\right) e^{-\frac{2\nu-3}{\nu-1/2}(N-N_*)}\,,
  \label{eq:inIRctPS}
\end{equation}
where $N-N_*$ measures the number of e-folds since the scale $k_*$ crossed the horizon and $dN = aH d\tau$. This formula for $\mathcal{P}_{v}$ is the same found for power-law inflation~\cite{Lyth:1991bc}, meaning these expressions are exact if $\eps$ is a true constant. In that case, inflation would continue forever. If $\nu > 3/2$, we easily see that $\mathcal{P}^{\rm reg}_{\zeta, h} \rightarrow \mathcal{P}^{}_{\zeta, h}$ in the IR, since $\mathcal{P}^{}_{\zeta, h}$ is time independent while the counterterm decays as $N \rightarrow \infty$. Interestingly, in the exact de Sitter case where $\eps = 0$ ($\nu = 3/2$), one can easily verify that $\mathcal{P}^{\rm reg}_v$ vanishes identically for all $(k,\tau)$, as pointed out by Parker~\cite{Parker:2007ni,Parker:2009uva}. 

On the other hand, if $\eps$ is only approximately constant, then inflation will end, which we take to occur at $\tau_0$.  CMB scales were supposed to have exited the horizon $N_{\rm CMB} = N_0 - N_* = 50-60$ e-folds before the end of inflation. Therefore if we evaluate the regularized power spectra at $\tau_0$, on CMB scales we obtain
\begin{align}
\mathcal{P}^{\rm reg}_{\zeta}(k_*,\tau_0) &\approx \frac{1}{8\pi^2\eps } \frac{H_{*}^2}{M_P^2} \bigg[ 1-e^{- (1-n_s)N_{\rm CMB} } + O(\varepsilon) \bigg] \,, \nonumber \\
\mathcal{P}^{\rm reg}_{h}(k_*,\tau_0) &\approx \frac{2}{\pi^2 } \frac{H_{*}^2}{M_P^2} \bigg[ 1-e^{-2 \varepsilon N_{\rm CMB} } + O(\varepsilon) \bigg] \,,
\end{align}
where $n_s -1 = 3-2\nu_s$ and we have expanded in the slow-roll limit while treating the argument of the exponential as $O(1)$. From the data on $n_s$ and $N_{\rm CMB}$~\cite{Planck:2018jri},  evaluation at the end of inflation would lead to corrections to the scalar spectrum of $O(10-25\%)$, while they can be much larger for the tensors. Indeed, it is clear that  $\mathcal{P}^{\rm reg}_{h}$ is $\eps$ suppressed as $\eps \rightarrow 0$. At first glance, this seems to confirm a problem as suggested in Refs.~\cite{Parker:2007ni,Agullo:2008ka,Agullo:2009vq,Agullo:2009zi}. However, because the counterterms continue to evolve, evaluating the renormalized power spectra at any point during inflation does not correspond to what is measured at late times in the CMB. To follow the evolution of the spectra further, we need to consider the transition to an FRW universe with a growing horizon at the end of inflation. 

\section{Spectrum of fluctuations after inflation} 
\label{sec:three}

Suppose we start in an inflating universe described by a constant equation of state $\win < -1/3$. At a time $\tau_0$, we then assume an instant transition to an FRW universe with a growing horizon described by another constant $\wout > -1/3$. We take the scale factor and the Hubble rate at the transition to be $a_0$ and $H_0$ respectively, and write $a(\tau)$ as
\begin{equation}
\frac{a(\tau)}{a_0} = 
\begin{cases} 
        (\tau/\tau_0)^{\frac{2}{(1+3\win)}} & \tau < \tau_0 \\
       \Big[ \frac{1}{2}a_0 H_0(\tau - \bar\tau)(1+3\wout)\Big]^{\frac{2}{(1+3\wout)}} & \tau > \tau_0\,,
   \end{cases}
\label{eq:SFmatching}
\end{equation}
where $\bar \tau/\tau_0=(w_2-w_1)/(w_2+1/3)$ and $a_0 H_0 \tau_0 = 2/(1+3w_1)$ are obtained by requiring continuity of $a$ and $a'$ at $\tau_0$. The solutions to~\cref{eq:MSeq} for these  ``\emph{in}" and ``\emph{out}" regions are given in terms of Hankel functions with indices $\nu$ and $\mu$, respectively. These indices are related with $w_1$ and $w_2$, as follows\footnote{For $|w| \leq 1$, these variables have the ranges $\nu \in [3/2, \infty)$ and $\mu \in [0,\infty)$.}
\begin{equation} \label{eq:winwout}
\win = \frac{3+2\nu}{3-6\nu} \,, \hspace{15mm} \wout = \frac{3-2\mu}{3+6\mu} \,.
\end{equation}
The solution for the modes in the inflating \emph{in} phase is given by~\cref{eq:BDmodes}
\begin{equation}
v_k^{\textrm{in}}(\tau) = \sqrt{\frac{\pi}{4k}}  e^{ i\frac{\pi}{4}(1+2\nu)} \sqrt{q} \, H_{\nu}^{(1)}(q) \equiv \frac{f_\nu (q)}{\sqrt{2k}}\,,
\end{equation}
while in the growing horizon \emph{out} phase we have
\begin{equation} \label{eq:OUTsol}
v_{k}^{\rm out}(\tau) =\frac{1}{\sqrt{2 k}} \Big( \alpha_k\, f_{\mu}(q-\bar q)+\beta_k \, f^*_{\mu}(q-\bar q) \Big) \, ,
\end{equation}
where $|\alpha_k|^2-|\beta_k|^2=1$ and $\bar q = -k\bar \tau$. The coefficients are determined by requiring the mode function and its derivative to be continuous at $\tau_0$, namely  $v_{k}^{\rm in}(\tau_0) = v_{k}^{\rm out}(\tau_0)$ and $v_{k}^{\prime\,\rm in}(\tau_0) = v_{k}^{\prime\,\rm out}(\tau_0)$. We find
\begin{align}
&\alpha_k = \frac{1}{2}\Big[f_{\nu}(q_0)f^*_{1+\mu}(\gamma q_0)+f_{\nu-1}(q_0)f^*_\mu(\gamma q_0)\Big]\, , \\
&\beta_k = \frac{1}{2}\Big[f_{\nu}(q_0)f_{1+\mu}(\gamma q_0)-f_{\nu-1}(q_0) f_\mu(\gamma q_0) \Big] \,,
\end{align}
where $\gamma=(1+2\mu)/(1-2\nu)$ and $q_0 = -k\tau_0$. 
Since $z''/z = a''/a$ in the \emph{out} region, the counterterm spectrum can be readily computed from~\cref{eq:ctSpec,eq:SFmatching}
\begin{equation}
  \mathcal{P}^{\rm ct}_{v}(k,\tau) = \frac{k^2}{4\pi^2} + \frac{(\mu^2-1/4)}{8\pi^2(\tau-\bar \tau)^2} \,.
\end{equation}
Let us first focus on the UV behavior of $\mathcal{P}_v(k,\tau)$ in the \emph{out} region. We find the following asymptotic form 
\begin{align}\label{eq:outUV}
\mathcal{P}_v(k,\tau)\, \xlongrightarrow[\text{UV}]{} \, &\frac{k^2}{4 \pi^2}+\frac{(\mu^2-1/4)}{8 \pi^2 (\tau-\bar \tau)^2} \\
& - \frac{(\mu+\nu) \cos(2 k(\tau-\tau_0))}{8 \pi^2 \gamma \tau_0^2}  + \mathcal{O}(k^{-2})\, . \notag
\end{align}
The first two UV divergent terms are exactly canceled by $\mathcal{P}^{\rm ct}_{v}$, while the oscillatory term is UV finite. In the IR limit, we find $\alpha_k\approx  \beta_k e^{-i \pi (\mu +\frac{1}{2})}$ and
\be
|\beta_k|^2\xlongrightarrow[\text{IR}]{}
\frac{4^{\nu+\mu}}{4\pi^2} \frac{q_0^{-2(\nu+\mu)}}{\gamma^{1+2 \mu}} \Gamma(\nu)^2 \Gamma(1+\mu)^2\, .
\ee
Using these relations, it can be shown that the unrenormalized power spectrum in the IR is {\it independent} of $\mu$. It reads\footnote{Helpful for deriving this expression is the identity $(q-\bar q)^{1+2\mu} = (a/a_0)^2(\gamma q_0)^{1+2\mu}$, which follows from~\cref{eq:SFmatching}. }
\begin{equation} \label{eq:psIRout}
\mathcal{P}_v(k,\tau)\, \xlongrightarrow[\text{IR}]{} \, \frac{a^2 H_{0}^2}{4\pi^2}\frac{2^{2\nu-3}\Gamma(\nu)^2}{\Gamma(3/2)^2}\left(\nu-\frac{1}{2}\right)^{1-2\nu} \left(\frac{k}{k_0} \right)^{3-2\nu}\, ,
\end{equation}
which is the same as~\cref{eq:inIRps} for a reference scale $k_0 = a_0 H_0$. The reference scale can be changed using $H_k^2 =H_0^2 (k/k_0)^{3-2\nu}$.  It is well-known that $\zeta$ and $h$ are expected to be conserved or ``frozen" on superhorizon scales. Here we see this explicitly in our computation by the fact that we obtain the same unregularized IR spectrum in the \emph{in} and \emph{out} regions, independent of the equation of state $w_2$ that describes the \emph{out} region.

We now turn to the fate of the IR-distorting counterterm. The counterterm spectrum in the IR is
\begin{equation}
\mathcal{P}^{\rm ct}_v(k,\tau)\, \xlongrightarrow[\text{IR}]{} \frac{a^2 H_0^2}{8\pi^2}\left(\frac{\mu - 1/2}{\mu+1/2}\right) e^{-\frac{(3+2\mu)}{(\mu+1/2)}(N-N_0)} \,  \, ,
\label{eq:outCTps}
\end{equation}
where $N-N_0$ measures the number of e-folds after the end of inflation. The factor in the exponential behaves as $(3+2\mu)/(\mu+1/2) \in (2,6]$ for $-1/3 < w_2 \leq 1$.  Recall that the $a^2$ growth for the Mukhanov variable cancels when one computes $\mathcal{P}^{\rm reg}_{\zeta,h}$. Then~\cref{eq:outCTps} shows that the IR counterterm spectrum for $\zeta$ and $h$ \emph{rapidly decays} after the end of inflation, while $\mathcal{P}_{\zeta,h}$ derived from~\cref{eq:psIRout} is independent of time. This means that $\mathcal{P}^{\rm reg}_{\zeta,h} \rightarrow \mathcal{P}_{\zeta,h}$ is an attractor solution in the IR that is reached a few e-folds after the end of inflation. Thus, we can safely take the limit $\tau \rightarrow \infty$ to obtain our final answer for the IR spectra of $\zeta$ and $h$ that would be measured at late times\footnote{Dividing by $\eps_{\rm in}$ when changing variables from $v_s$ to $\zeta$ is correct for the superhorizon spectrum, as it is frozen in during inflation. For more on the matching conditions for $\zeta$, see~\cite{Deruelle:1995kd,Dineen:2023nbt}.}
\begin{align}
\mathcal{P}^{\rm reg}_{\zeta}(k_*,\infty) &\approx \frac{1}{8\pi^2\eps_{\rm in} } \frac{H_{*}^2}{M_P^2} \equiv A_s \,,  \\
\mathcal{P}^{\rm reg}_{h}(k_*,\infty) &\approx \frac{2}{\pi^2 } \frac{H_{*}^2}{M_P^2} \equiv A_t\,,
\end{align}
which gives the standard prediction for the tensor-to-scalar ratio $r = A_t/A_s = 16\eps_{\rm in}$.

\begin{figure*}
  \centering
  \includegraphics[trim={0.6cm 0cm 0cm 0cm}, width=0.75\textwidth]{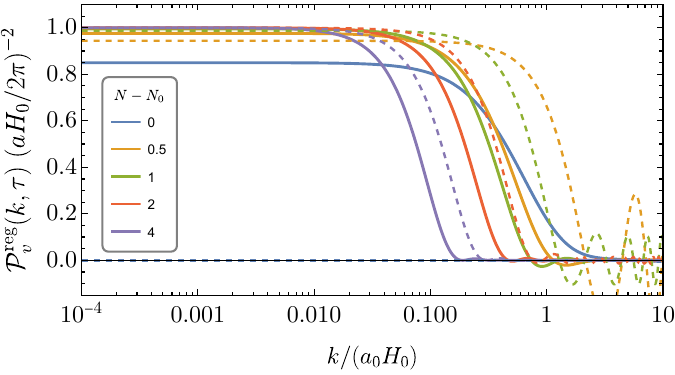}
  \caption{Regularized power spectrum for a universe that makes a transition from an inflating phase with $w_1 =-1$ (de Sitter) to a growing horizon phase with $w_2  = 0$ (matter). Solid lines are obtained by numerically solving~\cref{eq:MSeq} (for a transition on a timescale $H_0^{-1}$), while dashed lines give the instant transition approximation using~\cref{eq:OUTsol}. See text for more details.}
  \label{fig:PSplot}
\end{figure*}
The reader may wonder whether the instant transition accurately captures the relevant dynamics. To put these doubts to rest, we have also solved~\cref{eq:MSeq} numerically for a universe that transitions out of inflation in a finite time $H_0^{-1}$ (see~\cref{sec:numSol} for details) to a matter-dominated universe. The results compared to the instant transition approximation are shown in~\cref{fig:PSplot}. One sees that the instant transition leads to an over-excitation of UV modes, as observed by Ford~\cite{Ford:1986sy}. The scale-invariant UV oscillations appearing in the instant transition case are due to the finite term in~\cref{eq:outUV}, arising from interference when squaring~\cref{eq:OUTsol}. In any finite duration transition, one expects $\beta_k$ to be exponentially suppressed in the UV,\footnote{This behavior is necessary to ensure the finiteness of all higher $n$-point functions.} qualitatively as $e^{-k\Delta t}$, where $\Delta t$ is the duration of the transition. This predicts an exponential envelope for the UV oscillations, which is indeed observed in our numerical solution. 

On the other hand, we see that the instant transition provides an excellent approximation to the full numerical solution in the IR a few e-folds after inflation ends, which is the region of interest for cosmology. The reason is that \emph{any} transition is effectively instant for modes whose period of oscillation is much longer than the transition time, namely $k^{-1} \gg \Delta t$. Needless to say, the instant transition does not change the dynamics of IR modes that eventually re-enter the horizon after inflation. The model is therefore fully appropriate to describe CMB observables such as the inflationary spectra. 

Finally, we would like to point out that showing a matter-dominated transition in~\cref{fig:PSplot} is a conservative choice. This is because the scale-invariant counterterm would decay even faster for $w_2 > 0$, such as a radiation-dominated transition. In fact, it vanishes immediately for a radiation \emph{out} state in the instant transition approximation. This is a consequence of the fact that the modes have no logarithmic divergences in Minkowski space.

\section{Conclusions} 
\label{sec:conc}

In this letter, we have computed the renormalized spectra of scalar and tensor perturbations from inflation using the technique of adiabatic regularization. This is a prescription for deriving local counterterms that must be subtracted at all times and for all scales. We do this and follow the evolution of the renormalized spectra through the inflationary transition using an instant transition model supported by a full numerical solution. Our results are summarized in~\cref{fig:PSplot}, which shows that the standard result for the IR spectrum is recovered just a few e-folds after inflation ends, while the counterterms ensure that UV divergences are canceled at all times. Importantly, it does not matter if the power spectrum is defined via the bare two-point function or using the regularized quantity in the coincidence limit. The standard predictions for inflationary observables, based on the IR spectrum, are found in either case.

\section*{Acknowledgments}

We would like to thank Jose Navarro-Salas for useful comments on the manuscript, as well as Thomas Colas for helpful discussions concerning the purity of cosmological perturbations. The work of BAS is supported by the STFC (grant No. ST/X000753/1). SP is supported by the Leverhulme Trust (grant No. RPG-2021-299).

\appendix 

\section{Review of adiabatic regularization} 
\label{sec:ADR}
It is well known that the UV divergences encountered when doing quantum field theory in curved spacetime are worse than the usual Minkowskian ones. In general, these can be dealt with via generalizations of standard renormalization techniques, such as dimensional regularization or point-splitting~\cite{Birrell:1982ix}. However, these methods are impractical for numerical computations because one cannot numerically take the limits $d \rightarrow 4$ or $x'\rightarrow x$, respectively.

Due to this fact, we focus on the adiabatic regularization technique. It was first developed by Parker and Fulling~\cite{Parker:1974qw,Fulling:1974zr,Fulling:1974pu}, but is now a textbook method~\cite{Birrell:1982ix,Mukhanov:2007zz,Parker:2009uva}. Adiabatic regularization is a mode-by-mode subtraction scheme which is known to be equivalent to point-splitting or dimensional regularization~\cite{Birrell:1982ix}. Specifically, the adiabatic method is (at least in even dimensions) identical to DeWitt-Schwinger renormalization, and equivalent to the Hadamard renormalization up to the well-known renormalization ambiguities \cite{Pla:2022spt,delRio:2014bpa}. 
The subtraction terms are derived by examining the behavior of the mode functions in an adiabatically expanding universe, where particle number is an adiabatic invariant~\cite{1967PhDT75P,Parker:1969au}. The advantage of working directly with the modes is that an explicit cancellation of divergences can be had before any integrals over modes are performed. The resulting integral quantities, such as the two-point function, are then automatically finite without going through a procedure of adding counterterms to the Lagrangian. This makes it a powerful tool for practical cosmological applications in FRW spacetime~\cite{1978RSPSA.361..513B}. We remark that the adiabatic method is compatible with locality and general covariance.\footnote{It is worth mentioning that the following discussion will assume that the vacuum state $|0\rangle$  is Hadamard. This requirement specifies the singularity structure of the two-point function and ensures the existence of Wick polynomials of arbitrary order. For a detailed analysis see Refs. \cite{Brunetti:1999jn,hollands:2001fb,hollands:2001nf,hollands:2004yh}.  }

The starting point of the adiabatic expansion is the mode equation
\be \label{eq:eom-basic}
\varphi_k''+(\omega^2 + \sigma)\varphi_k=0\, .
\ee
For a massive scalar field in FRW $\phi$, the field modes $\varphi_k=\phi_k/a$ obey this equation with $\omega^2=k^2 + a^2 m^2$ and $\sigma=a^2(\xi-\frac{1}{6})R$ in conformal time. 
 The second step is to assume the Wentzel-Kramers-Brillouin (WKB) ansatz for the field modes 
\be
\varphi_k\sim \frac{1}{\sqrt{2\, \Omega_k(\tau)}}e^{-i\int^\tau \Omega_k(\tau') d\tau'}\, ,
\ee
where the function $\Omega_k$ is expanded adiabatically as
\be \label{eq:expansion}
\Omega_k=\sum_{n=0}^{\infty}\omega^{(n)}_k\, .
\ee
The super-index $^{(n)}$ refers to the adiabatic order of the expansion. Inserting the WKB ansatz in the mode equation we obtain 
\be
\Omega_k^2=\omega^2+\sigma+\frac{3}{4} \frac{{\Omega}_k^{\prime 2}}{\Omega_k^2}-\frac{1}{2} \frac{{\Omega}''_k}{\Omega_k} \,.
\ee
At this stage, it is crucial to properly fix the adiabatic order of the background: $\omega$ is a function of adiabatic order zero, while $\sigma$ is of adiabatic order two. Finally, inserting~\cref{eq:expansion} and grouping terms of the same adiabatic order we obtain, by iteration, the coefficients of the expansion. The first four orders are
\bea \label{eq:adiabatic:coefficients}
\omega^{(0)}&=&\omega\, ,\\
 \omega^{(1)}&=&\omega^{(3)}=0 \,,\\
 \omega^{(2)}&=&\frac{\sigma}{2\omega} +\frac{3}{8} \frac{(\omega')^2}{\omega^3}-\frac{1}{4} \frac{\omega''}{\omega^2}\, , \label{eq:adiabatic:O2}\\
 \omega^{(4)}&=&\frac{1}{2 \omega^3}\Big\{2 \sigma \omega \omega^{(2)}-5 \omega^2(\omega^{(2)})^2\\
 &&+\frac{3}{2} \omega' {\omega}^{\prime (2)}
 -\frac{1}{2}\left(\omega {\omega}^{\prime \prime (2)}+\omega^{(2)} \omega''\right)\Big\} \, .\nonumber
\eea
From the adiabatic expansion of the field modes, we can obtain the adiabatic expansion of composite quantities, such as the spectrum of the two-point function 
\be
2|\varphi_k(\tau)|^2_{\textrm{Ad}} \sim (\Omega^{-1}_k)^{(0)}+(\Omega^{-1}_k)^{(2)}+(\Omega^{-1}_k)^{(4)}...
\ee
As stated above, the adiabatic expansion precisely captures the UV behavior of the theory \cite{Parker:2009uva,Pirk:1992ye}. 
Therefore, it can be used to regularize observables since it removes their UV divergences by simply subtracting the adiabatic counterterms mode by mode (in the momentum space).\footnote{A modified version of adiabatic regularization was recently proposed accounting for the existence of ambiguities associated with the renormalization scale~\cite{Ferreiro:2022ibf,Ferreiro:2023uvr}.} The number of subtractions is determined by the scaling dimension of the observable.  In particular, the physical coincident two-point function reads
\be \label{eq:phiphys0}
\langle \phi^2\rangle _{\textrm{phys}}=\int\frac{dk}{k}\frac{k^3}{4\pi a^2}\left(2|\varphi_k|^2-(\Omega^{-1}_k)^{(0)}-(\Omega^{-1}_k)^{(2)}\right)\, .
\ee

This method can be easily applied to other cases as long as the equation for the field modes can be expressed as in~\cref{eq:eom-basic}. This is the case of the Mukhanov variable $v$ for both scalar and tensor perturbations, whose mode equation Eq. \eqref{eq:MSeq} has this form with $\omega=k$ and $\sigma=-z''/z$.
It implies that the coincident two-point function
\begin{equation}
\langle v^2 \rangle = \int_{0}^{\infty} \frac{dk}{k} \mathcal{P}_{v}(k,\tau)\,, \hspace{7mm} \mathcal{P}_{v}\equiv \frac{k^3}{2\pi^2} |v_k(\tau)|^2 \,,
\end{equation}
becomes UV finite after subtracting the two first adiabatic counterterms
\be
\mathcal{P}^{\rm reg}_v(k,\tau)=\mathcal{P}_v(k,\tau)-\mathcal{P}_v^{\textrm{ct}}(k,\tau)\, ,
\label{eq:regSpecAdiabatic}
\ee
where
\be \label{eq:PctAppendix}
\mathcal{P}_v^{\textrm{ct}}(k,\tau)=\frac{k^3}{2\pi^2}\left(\frac{1}{2\omega}-\frac{\omega^{(2)}}{2\omega^2}\right)=\frac{k^2}{4\pi^2}+\frac{1}{8\pi^2} \frac{z''}{z}\, ,
\ee
with $\omega^{(2)}$ given in Eq. \eqref{eq:adiabatic:O2}. We note that although the adiabatic expansion obeys the mode equation order-by-order, it is an asymptotic (UV) expansion that does not necessarily capture all the properties of the field modes. In particular, in an inflating universe, the power-spectrum for the Mukanhov variable $|v_k|^2$ the IR region grows as $a^2$ (for $\zeta,h$ the modes are frozen), while the adiabatic counterterms have in general a different behavior. For the first example considered in the main text, with $a\propto \tau^{1/2-\nu}$ and $\nu\geq 3/2$ we find $\mathcal{P}_v^{\textrm{ct}}\propto a^{\frac{2}{\nu-1/2}}$, which only coincides with the IR behavior of the modes in de Sitter.

\subsection{An infrared-safe definition for the counterterm spectrum}
\label{sec:IRsafePS}
Due to the scale-invariant term,~\cref{eq:PctAppendix} gives rise to a logarithmic IR divergence in $\langle v^2 \rangle$ when the massless limit $m\rightarrow 0$ is taken. Here we show that these IR divergences can be removed by an alternative, IR-safe definition for the counterterm spectrum. Similar conclusions about the absence of IR divergences were reached in~\cite{Negro:2024bbf}. Let us consider the ``resummed" spectrum~\cite{Ferreiro:2020uno}
\be
 \mathcal{\overline P}^{\textrm{ct}}(k,\tau)=\frac{k^3}{2\pi^2}\frac{1}{2\bar \omega} \, ,
 \label{eq:resummedPct}
\ee
where $\bar \omega^2=k^2 + a^2 m^2 + \sigma$. One can easily verify that $\mathcal{\overline P}^{\textrm{ct}}$ and $\mathcal{P}^{\textrm{ct}}$ differ in the UV only by finite terms 
\begin{equation}
\mathcal{P}^{\textrm{ct}} - \mathcal{\overline P}^{\textrm{ct}} \xlongrightarrow[\text{UV}]{} \frac{m^2 a a''}{16 \pi ^2 k^2}+\frac{m^2 (a')^2}{16 \pi ^2 k^2}-\frac{3 \sigma
   ^2}{32 \pi ^2 k^2} + O(k^{-4}) \,,
   \label{eq:PSfiniteShift}
\end{equation}
ensuring that $\mathcal{P}^{\rm reg}$ will remain UV finite. Unlike $\mathcal{P}^{\textrm{ct}}$, however, $ \mathcal{\overline P}^{\textrm{ct}}$ is fully IR finite. In fact, $ \mathcal{\overline P}^{\textrm{ct}}\rightarrow 0$ for $k^2 \ll a^2 m^2 + \sigma$. This is another way to recover our main result that $\mathcal{P}^{\rm reg}_v \rightarrow \mathcal{P}_v$ in the IR, i.e. that renormalization does not modify the superhorizon part of the spectrum.

Unlike the divergent terms, the UV finite behavior in~\cref{eq:PSfiniteShift} is not unique. For example, it may be changed by making a shift $\bar \omega^2 \rightarrow \bar \omega^2 + O(k^{-2})$. It is important to emphasize that when the resummed spectrum in~\cref{eq:resummedPct} is used, this shift has no effect in the IR and therefore cannot change the superhorizon spectrum. In the case of the massless Mukhanov variable, different choices for the $O(k^{-2})$ contribution lead to finite shifts in $\langle v^2 \rangle$ of the form
\begin{equation}
\langle v^2 \rangle_1 -\langle v^2 \rangle_2 = \frac{c}{16\pi^2} \frac{z''}{z}\,,
\end{equation}
where $c$ is a constant that can be changed by changing the renormalization scale. However, this ambiguity in $\langle v^2 \rangle$ disappears after the inflationary phase, where $z''/z$ acquires the exponential suppression shown in~\cref{eq:outCTps}.

\section{Details of the numerical solution} 
\label{sec:numSol}
We want to numerically solve the Mukhanov equation 
\begin{equation}
v_k''(\tau) + \left(k^2 - \frac{z''}{z} \right) v_k(\tau) = 0 \,.
\label{eq:MSeqNUM}
\end{equation}
For simplicity in describing our method, we restrict ourselves here to the case $z''/z = a''/a$ which describes the tensor spectrum (as we will see, the generalization to the scalars is straightforward).
Since $dN = H dt$, it is easier to work with real time when integrating over many e-folds of inflation where $H \approx$ constant and the integrator can take nearly constant time steps. The equivalent equation in real time $t$ and changing variables $v_k(t) \rightarrow u_k(t)/\sqrt{a(t)}$ is
\begin{equation}
\ddot{u}_k(t) + \left[\frac{k^2}{a^2} +\frac{1}{4} \left(3H^2 - R \right) \right] u_k(t) = 0 \,.
\label{eq:modeEQnum}
\end{equation}
The initial conditions for $u_k$ and $\dot{u}_k$ are 
\begin{align} \label{eq:BDmodesNUM}
u_k (k/aH\rightarrow \infty) & \rightarrow \sqrt{\frac{a}{2k}}  \exp \left(-ik \int^{t}\frac{dt'}{a(t')} \right) \,, \\
\dot{u}_k (k/aH\rightarrow \infty) &\rightarrow  u_k(t)  \left(\frac{H}{2} -\frac{ik}{a} \right) \,, \label{eq:BDmodesNUM2}
\end{align}
where $H = \dot{a}/a$. 
All we need now is to specify the background. This can be done by giving a continuous function for $\varepsilon(t) = -\dot{H}/H^2$. We choose the $C^\infty$ function
\begin{equation}
\eps(t) = \frac{\eps_+ }{2} \left[ 1+\frac{\eps_- }{\eps_+ }\tanh\left(\frac{t-t_0}{\Delta t}\right) \right] \,,
\end{equation}
where $\eps_+ = \eps_{\rm out} + \eps_{\rm in}$ and $\eps_- = \eps_{\rm out} - \eps_{\rm in}$.  This function describes a smooth transition from a universe with $\eps_{\rm in}$ (for $t\rightarrow -\infty$) to a universe described by $\eps_{\rm out}$ (for $t\rightarrow \infty$)  occurring at $t_0$ and on a timescale $\Delta t$. The relation between $\eps$ and the equation of state $w$ is $\eps = 3(1+w)/2$, so e.g. a transition from a de Sitter initial state to a radiation final state is described by $\eps_{\rm in} =0$ and $\eps_{\rm out} = 2$. This function completely specifies the background, i.e. it can be integrated to find $H$ and $a$. With these at hand, one can compute any background quantities such as $R$, $z$, $z''/z$, etc. Our setup therefore makes it simple to also solve~\cref{eq:MSeqNUM} for the scalar spectrum, where $z = a M_P \sqrt{2\eps}$.
By definition, inflation ends when $\varepsilon = 1$. 

Our numerical procedure is as follows.
We solve $n=512$ copies of ~\cref{eq:modeEQnum} by specifying $n$ values of $k$, with boundary conditions given by~\cref{eq:BDmodesNUM,eq:BDmodesNUM2}. Phase differences between modes are only physical when they interfere. Since here we are solving a linear theory, all modes evolve independently and the phase in~\cref{eq:BDmodesNUM} can be dropped. A boundary condition at infinity cannot be exactly applied in our finite numerical simulation. To this end, we define a penetration factor $P = (k/aH)_{\rm start}$ that determines when we apply the boundary conditions and begin integrating a particular mode. We find that the initial conditions are well approximated for $P >100$. To cover a range of scales of $k_{\rm UV}/k_{\rm IR} \approx 10^5$, we need to simulate $N \gtrsim \log(10^5) \approx 12$ e-folds of inflation. Taking the simulation to begin at $t=0$, the duration of inflation can be controlled by choosing $t_0$ appropriately, and we choose the duration of the transition to be $\Delta t = H_0^{-1}$. When computing the regularized power spectrum, we subtract the adiabatic counterterms on a mode-by-mode basis using the exact expression in~\cref{eq:PctAppendix}.

\bibliographystyle{JHEP}
\bibliography{refs.bib}

\providecommand{\href}[2]{#2}\begingroup\raggedright\begin{thebibliography}{10}

\bibitem{Guth:1980zm}
A.~H. Guth, {\it {The Inflationary Universe: A Possible Solution to the Horizon
  and Flatness Problems}},  {\em Phys. Rev. D} {\bf 23} (1981) 347--356.

\bibitem{Starobinsky:1980te}
A.~A. Starobinsky, {\it {A New Type of Isotropic Cosmological Models Without
  Singularity}},  {\em Phys. Lett. B} {\bf 91} (1980) 99--102.

\bibitem{Linde:1981mu}
A.~D. Linde, {\it {A New Inflationary Universe Scenario: A Possible Solution of
  the Horizon, Flatness, Homogeneity, Isotropy and Primordial Monopole
  Problems}},  {\em Phys. Lett. B} {\bf 108} (1982) 389--393.

\bibitem{Starobinsky:1979ty}
A.~A. Starobinsky, {\it {Spectrum of relict gravitational radiation and the
  early state of the universe}},  {\em JETP Lett.} {\bf 30} (1979) 682--685.

\bibitem{Mukhanov:1981xt}
V.~F. Mukhanov and G.~V. Chibisov, {\it {Quantum Fluctuations and a Nonsingular
  Universe}},  {\em JETP Lett.} {\bf 33} (1981) 532--535.

\bibitem{Hawking:1982cz}
S.~W. Hawking, {\it {The Development of Irregularities in a Single Bubble
  Inflationary Universe}},  {\em Phys. Lett. B} {\bf 115} (1982) 295.

\bibitem{Starobinsky:1982ee}
A.~A. Starobinsky, {\it {Dynamics of Phase Transition in the New Inflationary
  Universe Scenario and Generation of Perturbations}},  {\em Phys. Lett. B}
  {\bf 117} (1982) 175--178.

\bibitem{Guth:1982ec}
A.~H. Guth and S.~Y. Pi, {\it {Fluctuations in the New Inflationary Universe}},
   {\em Phys. Rev. Lett.} {\bf 49} (1982) 1110--1113.

\bibitem{Linde:1982uu}
A.~D. Linde, {\it {Scalar Field Fluctuations in Expanding Universe and the New
  Inflationary Universe Scenario}},  {\em Phys. Lett. B} {\bf 116} (1982)
  335--339.

\bibitem{Starobinsky:1983zz}
A.~A. Starobinsky, {\it {The Perturbation Spectrum Evolving from a Nonsingular
  Initially De-Sitter Cosmology and the Microwave Background Anisotropy}},
  {\em Sov. Astron. Lett.} {\bf 9} (1983) 302.

\bibitem{Bardeen:1983qw}
J.~M. Bardeen, P.~J. Steinhardt, and M.~S. Turner, {\it {Spontaneous Creation
  of Almost Scale - Free Density Perturbations in an Inflationary Universe}},
  {\em Phys. Rev. D} {\bf 28} (1983) 679.

\bibitem{Halliwell:1984eu}
J.~J. Halliwell and S.~W. Hawking, {\it {The Origin of Structure in the
  Universe}},  {\em Phys. Rev. D} {\bf 31} (1985) 1777.

\bibitem{COBE:1992syq}
{\bf COBE} Collaboration, G.~F. Smoot et~al., {\it {Structure in the COBE
  differential microwave radiometer first year maps}},  {\em Astrophys. J.
  Lett.} {\bf 396} (1992) L1--L5.

\bibitem{WMAP:2003elm}
{\bf WMAP} Collaboration, D.~N. Spergel et~al., {\it {First year Wilkinson
  Microwave Anisotropy Probe (WMAP) observations: Determination of cosmological
  parameters}},  {\em Astrophys. J. Suppl.} {\bf 148} (2003) 175--194,
  [\href{http://arxiv.org/abs/astro-ph/0302209}{{\tt astro-ph/0302209}}].

\bibitem{WMAP:2010qai}
{\bf WMAP} Collaboration, E.~Komatsu et~al., {\it {Seven-Year Wilkinson
  Microwave Anisotropy Probe (WMAP) Observations: Cosmological
  Interpretation}},  {\em Astrophys. J. Suppl.} {\bf 192} (2011) 18,
  [\href{http://arxiv.org/abs/1001.4538}{{\tt arXiv:1001.4538}}].

\bibitem{Planck:2013pxb}
{\bf Planck} Collaboration, P.~A.~R. Ade et~al., {\it {Planck 2013 results.
  XVI. Cosmological parameters}},  {\em Astron. Astrophys.} {\bf 571} (2014)
  A16, [\href{http://arxiv.org/abs/1303.5076}{{\tt arXiv:1303.5076}}].

\bibitem{Planck:2015fie}
{\bf Planck} Collaboration, P.~A.~R. Ade et~al., {\it {Planck 2015 results.
  XIII. Cosmological parameters}},  {\em Astron. Astrophys.} {\bf 594} (2016)
  A13, [\href{http://arxiv.org/abs/1502.01589}{{\tt arXiv:1502.01589}}].

\bibitem{Planck:2018vyg}
{\bf Planck} Collaboration, N.~Aghanim et~al., {\it {Planck 2018 results. VI.
  Cosmological parameters}},  {\em Astron. Astrophys.} {\bf 641} (2020) A6,
  [\href{http://arxiv.org/abs/1807.06209}{{\tt arXiv:1807.06209}}]. [Erratum:
  Astron.Astrophys. 652, C4 (2021)].

\bibitem{BICEP:2021xfz}
{\bf BICEP, Keck} Collaboration, P.~A.~R. Ade et~al., {\it {Improved
  Constraints on Primordial Gravitational Waves using Planck, WMAP, and
  BICEP/Keck Observations through the 2018 Observing Season}},  {\em Phys. Rev.
  Lett.} {\bf 127} (2021), no.~15 151301,
  [\href{http://arxiv.org/abs/2110.00483}{{\tt arXiv:2110.00483}}].

\bibitem{Planck:2018jri}
{\bf Planck} Collaboration, Y.~Akrami et~al., {\it {Planck 2018 results. X.
  Constraints on inflation}},  {\em Astron. Astrophys.} {\bf 641} (2020) A10,
  [\href{http://arxiv.org/abs/1807.06211}{{\tt arXiv:1807.06211}}].

\bibitem{Bezrukov:2007ep}
F.~L. Bezrukov and M.~Shaposhnikov, {\it {The Standard Model Higgs boson as the
  inflaton}},  {\em Phys. Lett. B} {\bf 659} (2008) 703--706,
  [\href{http://arxiv.org/abs/0710.3755}{{\tt arXiv:0710.3755}}].

\bibitem{Kallosh:2013hoa}
R.~Kallosh and A.~Linde, {\it {Universality Class in Conformal Inflation}},
  {\em JCAP} {\bf 07} (2013) 002, [\href{http://arxiv.org/abs/1306.5220}{{\tt
  arXiv:1306.5220}}].

\bibitem{Kallosh:2013daa}
R.~Kallosh and A.~Linde, {\it {Multi-field Conformal Cosmological Attractors}},
   {\em JCAP} {\bf 12} (2013) 006, [\href{http://arxiv.org/abs/1309.2015}{{\tt
  arXiv:1309.2015}}].

\bibitem{Ferrara:2013rsa}
S.~Ferrara, R.~Kallosh, A.~Linde, and M.~Porrati, {\it {Minimal Supergravity
  Models of Inflation}},  {\em Phys. Rev. D} {\bf 88} (2013), no.~8 085038,
  [\href{http://arxiv.org/abs/1307.7696}{{\tt arXiv:1307.7696}}].

\bibitem{Farakos:2013cqa}
F.~Farakos, A.~Kehagias, and A.~Riotto, {\it {On the Starobinsky Model of
  Inflation from Supergravity}},  {\em Nucl. Phys. B} {\bf 876} (2013)
  187--200, [\href{http://arxiv.org/abs/1307.1137}{{\tt arXiv:1307.1137}}].

\bibitem{Mukhanov:2013tua}
V.~Mukhanov, {\it {Quantum Cosmological Perturbations: Predictions and
  Observations}},  {\em Eur. Phys. J. C} {\bf 73} (2013) 2486,
  [\href{http://arxiv.org/abs/1303.3925}{{\tt arXiv:1303.3925}}].

\bibitem{CMB-S4:2016ple}
{\bf CMB-S4} Collaboration, K.~N. Abazajian et~al., {\it {CMB-S4 Science Book,
  First Edition}},  \href{http://arxiv.org/abs/1610.02743}{{\tt
  arXiv:1610.02743}}.

\bibitem{Abazajian:2019eic}
K.~Abazajian et~al., {\it {CMB-S4 Science Case, Reference Design, and Project
  Plan}},  \href{http://arxiv.org/abs/1907.04473}{{\tt arXiv:1907.04473}}.

\bibitem{SimonsObservatory:2018koc}
{\bf Simons Observatory} Collaboration, P.~Ade et~al., {\it {The Simons
  Observatory: Science goals and forecasts}},  {\em JCAP} {\bf 02} (2019) 056,
  [\href{http://arxiv.org/abs/1808.07445}{{\tt arXiv:1808.07445}}].

\bibitem{Parker:2007ni}
L.~Parker, {\it {Amplitude of Perturbations from Inflation}},
  \href{http://arxiv.org/abs/hep-th/0702216}{{\tt hep-th/0702216}}.

\bibitem{Agullo:2008ka}
I.~Agullo, J.~Navarro-Salas, G.~J. Olmo, and L.~Parker, {\it {Reexamination of
  the Power Spectrum in De Sitter Inflation}},  {\em Phys. Rev. Lett.} {\bf
  101} (2008) 171301, [\href{http://arxiv.org/abs/0806.0034}{{\tt
  arXiv:0806.0034}}].

\bibitem{Agullo:2009vq}
I.~Agullo, J.~Navarro-Salas, G.~J. Olmo, and L.~Parker, {\it {Revising the
  predictions of inflation for the cosmic microwave background anisotropies}},
  {\em Phys. Rev. Lett.} {\bf 103} (2009) 061301,
  [\href{http://arxiv.org/abs/0901.0439}{{\tt arXiv:0901.0439}}].

\bibitem{Agullo:2009zi}
I.~Agullo, J.~Navarro-Salas, G.~J. Olmo, and L.~Parker, {\it {Revising the
  observable consequences of slow-roll inflation}},  {\em Phys. Rev. D} {\bf
  81} (2010) 043514, [\href{http://arxiv.org/abs/0911.0961}{{\tt
  arXiv:0911.0961}}].

\bibitem{delRio:2014aua}
A.~del Rio and J.~Navarro-Salas, {\it {Spacetime correlators of perturbations
  in slow-roll de Sitter inflation}},  {\em Phys. Rev. D} {\bf 89} (2014),
  no.~8 084037, [\href{http://arxiv.org/abs/1401.6912}{{\tt arXiv:1401.6912}}].

\bibitem{Urakawa:2009xaa}
Y.~Urakawa and A.~A. Starobinsky, {\it {Adiabatic regularization of primordial
  perturbations generated during inflation}},  in {\em {19th Workshop on
  General Relativity and Gravitation in Japan}}, 2009.

\bibitem{Choudhury:2023rks}
S.~Choudhury, S.~Panda, and M.~Sami, {\it {Quantum loop effects on the power
  spectrum and constraints on primordial black holes}},  {\em JCAP} {\bf 11}
  (2023) 066, [\href{http://arxiv.org/abs/2303.06066}{{\tt arXiv:2303.06066}}].

\bibitem{Parker:1974qw}
L.~Parker and S.~A. Fulling, {\it {Adiabatic regularization of the energy
  momentum tensor of a quantized field in homogeneous spaces}},  {\em Phys.
  Rev. D} {\bf 9} (1974) 341--354.

\bibitem{Fulling:1974zr}
S.~A. Fulling and L.~Parker, {\it {Renormalization in the theory of a quantized
  scalar field interacting with a robertson-walker spacetime}},  {\em Annals
  Phys.} {\bf 87} (1974) 176--204.

\bibitem{Fulling:1974pu}
S.~A. Fulling, L.~Parker, and B.~L. Hu, {\it {Conformal energy-momentum tensor
  in curved spacetime: Adiabatic regularization and renormalization}},  {\em
  Phys. Rev. D} {\bf 10} (1974) 3905--3924.

\bibitem{Birrell:1982ix}
N.~D. Birrell and P.~C.~W. Davies, {\em {Quantum Fields in Curved Space}}.
\newblock Cambridge Monographs on Mathematical Physics. Cambridge Univ. Press,
  Cambridge, UK, 2, 1984.

\bibitem{Mukhanov:2007zz}
V.~Mukhanov and S.~Winitzki, {\em {Introduction to quantum effects in
  gravity}}.
\newblock Cambridge University Press, 6, 2007.

\bibitem{Parker:2009uva}
L.~E. Parker and D.~Toms, {\em {Quantum Field Theory in Curved Spacetime}:
  {Quantized Field and Gravity}}.
\newblock Cambridge Monographs on Mathematical Physics. Cambridge University
  Press, 8, 2009.

\bibitem{Durrer:2009ii}
R.~Durrer, G.~Marozzi, and M.~Rinaldi, {\it {On Adiabatic Renormalization of
  Inflationary Perturbations}},  {\em Phys. Rev. D} {\bf 80} (2009) 065024,
  [\href{http://arxiv.org/abs/0906.4772}{{\tt arXiv:0906.4772}}].

\bibitem{Marozzi:2011da}
G.~Marozzi, M.~Rinaldi, and R.~Durrer, {\it {On infrared and ultraviolet
  divergences of cosmological perturbations}},  {\em Phys. Rev. D} {\bf 83}
  (2011) 105017, [\href{http://arxiv.org/abs/1102.2206}{{\tt
  arXiv:1102.2206}}].

\bibitem{Agullo:2011qg}
I.~Agullo, J.~Navarro-Salas, G.~J. Olmo, and L.~Parker, {\it {Remarks on the
  renormalization of primordial cosmological perturbations}},  {\em Phys. Rev.
  D} {\bf 84} (2011) 107304, [\href{http://arxiv.org/abs/1108.0949}{{\tt
  arXiv:1108.0949}}].

\bibitem{Bastero-Gil:2013nja}
M.~Bastero-Gil, A.~Berera, N.~Mahajan, and R.~Rangarajan, {\it {Power spectrum
  generated during inflation}},  {\em Phys. Rev. D} {\bf 87} (2013), no.~8
  087302, [\href{http://arxiv.org/abs/1302.2995}{{\tt arXiv:1302.2995}}].

\bibitem{Wang:2015zfa}
D.-G. Wang, Y.~Zhang, and J.-W. Chen, {\it {Vacuum and gravitons of relic
  gravitational waves and the regularization of the spectrum and
  energy-momentum tensor}},  {\em Phys. Rev. D} {\bf 94} (2016), no.~4 044033,
  [\href{http://arxiv.org/abs/1512.03134}{{\tt arXiv:1512.03134}}].

\bibitem{Markkanen:2017rvi}
T.~Markkanen, {\it {Renormalization of the inflationary perturbations
  revisited}},  {\em JCAP} {\bf 05} (2018) 001,
  [\href{http://arxiv.org/abs/1712.02372}{{\tt arXiv:1712.02372}}].

\bibitem{Corba:2022ugu}
S.~P. Corb\`a and L.~Sorbo, {\it {On adiabatic subtraction in an inflating
  Universe}},  {\em JCAP} {\bf 07} (2023) 005,
  [\href{http://arxiv.org/abs/2209.14362}{{\tt arXiv:2209.14362}}].

\bibitem{Ferreiro:2022ibf}
A.~Ferreiro and F.~Torrenti, {\it {Ultraviolet-regularized power spectrum
  without infrared distortions in cosmological spacetimes}},  {\em Phys. Lett.
  B} {\bf 840} (2023) 137868, [\href{http://arxiv.org/abs/2212.01078}{{\tt
  arXiv:2212.01078}}].

\bibitem{Ferreiro:2023uvr}
A.~Ferreiro, S.~Monin, and F.~Torrenti, {\it {Physical scale adiabatic
  regularization in cosmological spacetimes}},
  \href{http://arxiv.org/abs/2311.08986}{{\tt arXiv:2311.08986}}.

\bibitem{Finelli:2007fr}
F.~Finelli, G.~Marozzi, G.~P. Vacca, and G.~Venturi, {\it {The Impact of
  ultraviolet regularization on the spectrum of curvature perturbations during
  inflation}},  {\em Phys. Rev. D} {\bf 76} (2007) 103528,
  [\href{http://arxiv.org/abs/0707.1416}{{\tt arXiv:0707.1416}}].

\bibitem{Woodard:2017zfq}
R.~P. Woodard, {\em {Perturbative quantum gravity comes of age}}, vol.~Volume
  2, pp.~349--414.
\newblock 2017.

\bibitem{Polarski:1995jg}
D.~Polarski and A.~A. Starobinsky, {\it {Semiclassicality and decoherence of
  cosmological perturbations}},  {\em Class. Quant. Grav.} {\bf 13} (1996)
  377--392, [\href{http://arxiv.org/abs/gr-qc/9504030}{{\tt gr-qc/9504030}}].

\bibitem{Mukhanov:1990me}
V.~F. Mukhanov, H.~A. Feldman, and R.~H. Brandenberger, {\it {Theory of
  cosmological perturbations. Part 1. Classical perturbations. Part 2. Quantum
  theory of perturbations. Part 3. Extensions}},  {\em Phys. Rept.} {\bf 215}
  (1992) 203--333.

\bibitem{Bardeen:1980kt}
J.~M. Bardeen, {\it {Gauge Invariant Cosmological Perturbations}},  {\em Phys.
  Rev. D} {\bf 22} (1980) 1882--1905.

\bibitem{Mukhanov:1988jd}
V.~F. Mukhanov, {\it {Quantum Theory of Gauge Invariant Cosmological
  Perturbations}},  {\em Sov. Phys. JETP} {\bf 67} (1988) 1297--1302.

\bibitem{Mukhanov:1989rq}
V.~F. Mukhanov, {\it {Quantum Theory of Cosmological Perturbations in R(2)
  Gravity}},  {\em Phys. Lett. B} {\bf 218} (1989) 17--20.

\bibitem{Misner:1973prb}
C.~W. Misner, K.~S. Thorne, and J.~A. Wheeler, {\em {Gravitation}}.
\newblock W. H. Freeman, San Francisco, 1973.

\bibitem{Lidsey:1995np}
J.~E. Lidsey, A.~R. Liddle, E.~W. Kolb, E.~J. Copeland, T.~Barreiro, and
  M.~Abney, {\it {Reconstructing the inflation potential : An overview}},  {\em
  Rev. Mod. Phys.} {\bf 69} (1997) 373--410,
  [\href{http://arxiv.org/abs/astro-ph/9508078}{{\tt astro-ph/9508078}}].

\bibitem{Maldacena:2002vr}
J.~M. Maldacena, {\it {Non-Gaussian features of primordial fluctuations in
  single field inflationary models}},  {\em JHEP} {\bf 05} (2003) 013,
  [\href{http://arxiv.org/abs/astro-ph/0210603}{{\tt astro-ph/0210603}}].

\bibitem{Bunch:1978yq}
T.~S. Bunch and P.~C.~W. Davies, {\it {Quantum Field Theory in de Sitter Space:
  Renormalization by Point Splitting}},  {\em Proc. Roy. Soc. Lond. A} {\bf
  360} (1978) 117--134.

\bibitem{Lyth:1991bc}
D.~H. Lyth and E.~D. Stewart, {\it {The Curvature perturbation in power law
  (e.g. extended) inflation}},  {\em Phys. Lett. B} {\bf 274} (1992) 168--172.

\bibitem{Deruelle:1995kd}
N.~Deruelle and V.~F. Mukhanov, {\it {On matching conditions for cosmological
  perturbations}},  {\em Phys. Rev. D} {\bf 52} (1995) 5549--5555,
  [\href{http://arxiv.org/abs/gr-qc/9503050}{{\tt gr-qc/9503050}}].

\bibitem{Dineen:2023nbt}
D.~D. Dineen and W.~J. Handley, {\it {Analytic Approximations for the
  Primordial Power Spectrum with Israel Junction Conditions}},
  \href{http://arxiv.org/abs/2309.15984}{{\tt arXiv:2309.15984}}.

\bibitem{Ford:1986sy}
L.~H. Ford, {\it {Gravitational Particle Creation and Inflation}},  {\em Phys.
  Rev. D} {\bf 35} (1987) 2955.

\bibitem{Pla:2022spt}
S.~Pla and E.~Winstanley, {\it {Equivalence of the adiabatic expansion and
  Hadamard renormalization for a charged scalar field}},  {\em Phys. Rev. D}
  {\bf 107} (2023), no.~2 025004, [\href{http://arxiv.org/abs/2209.01079}{{\tt
  arXiv:2209.01079}}].

\bibitem{delRio:2014bpa}
A.~del Rio and J.~Navarro-Salas, {\it {Equivalence of Adiabatic and
  DeWitt-Schwinger renormalization schemes}},  {\em Phys. Rev. D} {\bf 91}
  (2015) 064031, [\href{http://arxiv.org/abs/1412.7570}{{\tt
  arXiv:1412.7570}}].

\bibitem{1967PhDT75P}
L.~E. {Parker}, {\em {The Creation of Particles in an Expanding Universe.}}
\newblock PhD thesis, Harvard University, Massachusetts, Jan., 1967.

\bibitem{Parker:1969au}
L.~Parker, {\it {Quantized fields and particle creation in expanding universes.
  1.}},  {\em Phys. Rev.} {\bf 183} (1969) 1057--1068.

\bibitem{1978RSPSA.361..513B}
N.~D. {Birrell}, {\it {The Application of Adiabatic Regularization to
  Calculations of Cosmological Interest}},  {\em Proceedings of the Royal
  Society of London Series A} {\bf 361} (June, 1978) 513--526.

\bibitem{Brunetti:1999jn}
R.~Brunetti and K.~Fredenhagen, {\it {Microlocal analysis and interacting
  quantum field theories: Renormalization on physical backgrounds}},  {\em
  Commun. Math. Phys.} {\bf 208} (2000) 623--661,
  [\href{http://arxiv.org/abs/math-ph/9903028}{{\tt math-ph/9903028}}].

\bibitem{hollands:2001fb}
S.~Hollands and R.~M. Wald, {\it {Existence of local covariant time ordered
  products of quantum fields in curved space-time}},  {\em Commun. Math. Phys.}
  {\bf 231} (2002) 309--345, [\href{http://arxiv.org/abs/gr-qc/0111108}{{\tt
  gr-qc/0111108}}].

\bibitem{hollands:2001nf}
S.~Hollands and R.~M. Wald, {\it {Local Wick polynomials and time ordered
  products of quantum fields in curved space-time}},  {\em Commun. Math. Phys.}
  {\bf 223} (2001) 289--326, [\href{http://arxiv.org/abs/gr-qc/0103074}{{\tt
  gr-qc/0103074}}].

\bibitem{hollands:2004yh}
S.~Hollands and R.~M. Wald, {\it {Conservation of the stress tensor in
  interacting quantum field theory in curved spacetimes}},  {\em Rev. Math.
  Phys.} {\bf 17} (2005) 227--312,
  [\href{http://arxiv.org/abs/gr-qc/0404074}{{\tt gr-qc/0404074}}].

\bibitem{Pirk:1992ye}
K.-T. Pirk, {\it {Hadamard states and adiabatic vacua}},  {\em Phys. Rev. D}
  {\bf 48} (1993) 3779--3783, [\href{http://arxiv.org/abs/gr-qc/9211003}{{\tt
  gr-qc/9211003}}].

\bibitem{Negro:2024bbf}
A.~Negro and S.~P. Patil, {\it {An \'Etude on the regularization and
  renormalization of divergences in primordial observables}},  {\em Riv. Nuovo
  Cim.} {\bf 47} (2024), no.~3 179--228,
  [\href{http://arxiv.org/abs/2402.10008}{{\tt arXiv:2402.10008}}].

\bibitem{Ferreiro:2020uno}
A.~Ferreiro, J.~Navarro-Salas, and S.~Pla, {\it {R-summed form of adiabatic
  expansions in curved spacetime}},  {\em Phys. Rev. D} {\bf 101} (2020),
  no.~10 105011, [\href{http://arxiv.org/abs/2003.09610}{{\tt
  arXiv:2003.09610}}].

\end{thebibliography}\endgroup

\end{document}